\documentclass[twocolumn,footinbib,showpacs,showkeys,superscriptaddress] {revtex4-1}
\usepackage{epsfig}
\usepackage{graphicx,graphics}
\usepackage{epstopdf}
\usepackage{natbib}
\usepackage{lipsum}
\usepackage{dcolumn}
\usepackage{amsmath,amssymb,amsfonts}
\usepackage{latexsym,verbatim}
\usepackage{bm}
\usepackage{color}
\usepackage{siunitx}
\usepackage[breaklinks=true,colorlinks,citecolor=blue,linkcolor=blue,urlcolor=blue]{hyperref}
\usepackage[utf8]{inputenc}

\begin{document}
\preprint{AIP/123-QED}	
\title{Spin Birefringence and Spin Polarized Transmission across p-wave Altermagnetic Heterostructure}
\author{Yong-Mei Zhang}
 \homepage{zymzym@nuaa.edu.cn}
\author{Zhi-Ping Niu}
\affiliation{College of Physics, Nanjing University of Aeronautics and Astronautics, Jiangsu 210016, People's Republic of China}

\date{\today}

\begin{abstract}
This paper investigates electron transport properties across N/AM interface or N/AM/N heterojunction. Based on Hamiltonians in the three regions, general wave functions are proposed and combined linear equations are constructed. Reflection and transmission coefficients as well as spin polarization of the altermagnet heterostructure are obtained. When electron incident from normal metal to altermagnet, it is scattered and split into $\prime+\prime$ branch and $\prime-\prime$ branch transmitting into altermagnet zone with separate refractive angles. Total internal reflection might happen to $\prime+\prime$ branch. Critical angle of total reflection is determined by incident conditions and properties of AM material. In the heterostructure case, spin-up and spin-down electrons transmitted with same propagation directions but different transmission probabilities. Transmittance and polarization periodically change with length of the junction. Exchange field and spin-splitting strength have pronounced effect on spin-polarized transmittance while the effect of spin-orbit coupling is quite small. Optimized system can be found by tuning parameters of the AM material. This study will lay the theoretical groundwork for experimental research on spin-selective outputs and spintronic devices.

\end{abstract}


\maketitle

\section {INTRODUCTION}
The concept of spin birefringence arose from analogous of optic birefringence. It describes electrons in condensed matter experience different effective refractive index and transmittance due to opposite spin states\cite{REF1,REF2}. It firstly drew attention in semi-conductor spintronics and got intensive investigation in graphene system\cite{REF3,REF4}. In inhomogeneous regions, spin-orbit coupling lifts spin degeneracy of Dirac fermion. Electrons with different spin orientations exhibit a splitting behavior at the interface analogous to that of light waves passing through a birefringent crystal, which is known as the spin-birefringence effect \cite{REF4, REF5}. This mechanism provides possibilities for all electronic controlling of spin without external magnetic field\cite{REF6, REF7}. It underlies significant physics fundamentals of designing spintronic devices.

Recently, along with deep investigation of spin transport properties in graphene, people began to explore spin birefringence in other 2D and 3D systems\cite{Bhattacharyya2023}. However, traditional realization of spin birefringence depends on strong spin-orbit coupling\cite{REF8,REF9}. This confines range of candidate materials and brings about adverse factors such as spin dephasing. The important investigation direction of spin birefringence is to find new material system to realize high efficiency spin birefringence without spin-orbit coupling.

The emergence of altermagnets(AM) provides over new perspective for this tough problem\cite{Smejkal031042,Smejkal040501,Feng2022}. Altermagnet is a new type of magnetic material recently being theoretically predicted and verified gradually in experiments \cite{Smejkal031042,Smejkal040501,REF10,REF11}. Its essential characteristics can be summarized in two respects. On the one hand, it has zero magnetic moment, which is similar to antiferromagnets. This circumvents the stray-field issue that is intrinsic to ferromagnets. On the other hand, altermagnet exhibits anisotropic spin splitting in momentum space\cite{REF12, REF13}. This kind of spin splitting arises from lattice potential, instead of spin-orbit coupling. The strength of spin-splitting can be one order of magnitude higher than that of the conventional spin orbit coupling effect \cite{REF11}. This non-relativistic spin splitting ensures altermagnet great potential in spintronics. Up to now, DFT calculations have predicted hundreds of altermagnets. Experimental methods like Angle-Resolved Photoemission Spectroscopy (ARPES) have also provided preliminary verification \cite{REF13}.

In the background of fast promotion of research of altermagnets, a natural and essential problem is whether this spin splitting, which is driven by lattice symmetry rather than spin orbit coupling, can induce unique spin-birefringent behavior. This question has multiple meanings. From the perspective of fundamental physics, the spin splitting in altermagnets is momentum dependent anisotropic. For example, in $d$-wave altermagnet, momentum dependent spin splitting is expressed as $E_{\pm}(\mathbf{k}) = tk^2 \pm J(k_x^2 - k_y^2)$ \cite{REF10,REF11,Bhowal2024}. Researchers may wonder whether the anisotropy in altermagnet generates entirely different scattering behavior apart from graphene when electrons come across interface. From the device application aspect, when birefringence in altermagnet is verified and can be tuned, it will offer new all electronic approaches to realize spin splitter, spin diode and other spintronic devices. It is revealed in recent research work that intrinsic spin processing in altermagnets could induce periodically oscillation in spatial spin distribution \cite{REF14}. Moreover, the oscillation period is determined by the strength of spin splitting. This supplies theoretical basis for measurement of parameters.

Currently, the research hotspot of altermagnets focus lies in several aspects. Firstly, material prediction and experimental validation continue to be essential endeavors. Among the recent significant progresses is the design and selection of two-dimensional monolayer altermagnets. By employing high-throughput first-principles calculations, researchers have successfully identified hundreds of potential two-dimensional altermagnets from thousands of candidate materials \cite{REF15}. Secondly, unconventional transport properties of altermagnets have attracted widespread attention, including spin-orbit coupling anomalous Hall effect, spin filter effect and valley-split regulation induced by spin-layer locking. Specifically, recent theoretical work reveals the possibility of consistent spin superfluid in the super conductive state of altermagnet \cite{REF10,REF16}, since spin-up and spin-down electrons form two independent superconducting condensates, they can carry a pure spin supercurrent with zero net charge current, and this behavior can persist even in the presence of spin-orbit coupling \cite{REF17,REF18}. These progresses point to the central conclusion that spin-splitting in altermagnets provides a profound and peculiar platform for exploring new phenomenon of spin transport. Spin birefringence is the right point of penetration connecting energy band features and device functions.

It is methodologically clear in its foundations to investigate spin birefringence in altermagnets. The study of birefringence adopts a typical research paradigm. That is to compose the system with normal region and spin-splitting altermagnet region so as to study the transmittance, refraction and spin polarization at the interface \cite{REF4,Bhattacharyya2023}. In altermagnet region, the strength of spin-splitting can be controlled by external electric field. The height of barrier is spin selective. This constitutes natural advantages for constructing all electric tuned spintronics devices \cite{REF10}. Based on lattice model, quantum transport calculation and Green's functions methods have been successfully applied to research of altermagnet properties\cite{REF19,REF20}. Spin-resolved transmittance, oscillation of Hall voltage and other observable signals can be described quantitatively. In addition, linear magneto-birefringence (LMB) has been put forward as an optic measurement to detect altermagnet order\cite{REF11}. Deep symmetric correlation might exist between LMB and spin birefringence. A comparative study of the two contributes to the construction of a unified conceptual framework \cite{REF11}.

Research on spin birefringence in altermagnets is not only driven by a clear physical motivation and supported by a solid theoretical framework, but also highly consistent with current hotspot in condensed matter physics and spintronics \cite{REF11,Bhattacharyya2023}. It is likely to reveal the unique contribution of nonrelativistic spin-splitting to spin birefringence by investigating spin-resolved transport properties at the interface or heterojunction of altermagnets. It will also provide theoretical support and material foundation for the new generation of low power consumption and nonmagnetic spintronic devices \cite{REF22}.

In the study of altermagnets, $p$-wave altermagnets exhibit several notable advantages over other types such as d-wave, particularly in their unique transport behavior in superconducting Josephson junctions and their spin transport properties \cite{10.1039/d6mh00357e}. According to current research in condensed matter physics, there are primarily two distinct pathways for the formation of $p$-wave altermagnets: one is through the spontaneous formation of non-collinear magnetic structures, and the other is through external manipulation (such as optical illumination) induced in specific lattice systems. In spintronics, $p$-wave magnets possess a distinctive tunneling magnetoresistance (TMR) effect. Unlike $d$-wave altermagnets, in a junction composed of two $p$-wave magnets, the relative rotation angle required to achieve the maximum TMR is $\pi$ ($180^{\circ}$), rather than $\pi/2$ ($90^{\circ}$). This provides a new degree of freedom for the design and manipulation of spin-valve devices \cite{2024Minimal} . Furthermore, $p$-wave magnets can generate equilibrium spin currents in the absence of an external electric field, as well as non-equilibrium spin currents under an applied electric field, making them promising candidates for spintronic device applications \cite{Leon2025,Niu2024}.

In this work, we investigate spin birefringence in heterostructures composed of AMs and normal metals. The theoretical approach is developed, starting with the altermagnet Hamiltonian and eigenfunctions in the heterojunction. By the use of continuous conditions on the boundaries, combined linear equations are derived, with which spin birefringence and spin resolved transmittance are obtained. Adjusting geometric and/or physical parameters, birefringence and spin-polarized transport are calculated numerically. Scattering at N/AM interface demonstrates spin birefringence tuned by electron energy, incident angle and AM material parameters. Transmission through the AM heterostructure exhibits periodic oscillation with length and spin polarization. This helps to establish all-electrical spintronic functionality that harnesses the intrinsic, non-relativistic spin splitting of AMs.

\section {Effective Hamiltonian and Spin Resolved Transmission of Altermagnet Heterosturcture}
\subsection{Polarized Transmission of N/AM/N Structure}

\begin{figure}
	\includegraphics[width=0.9\columnwidth]{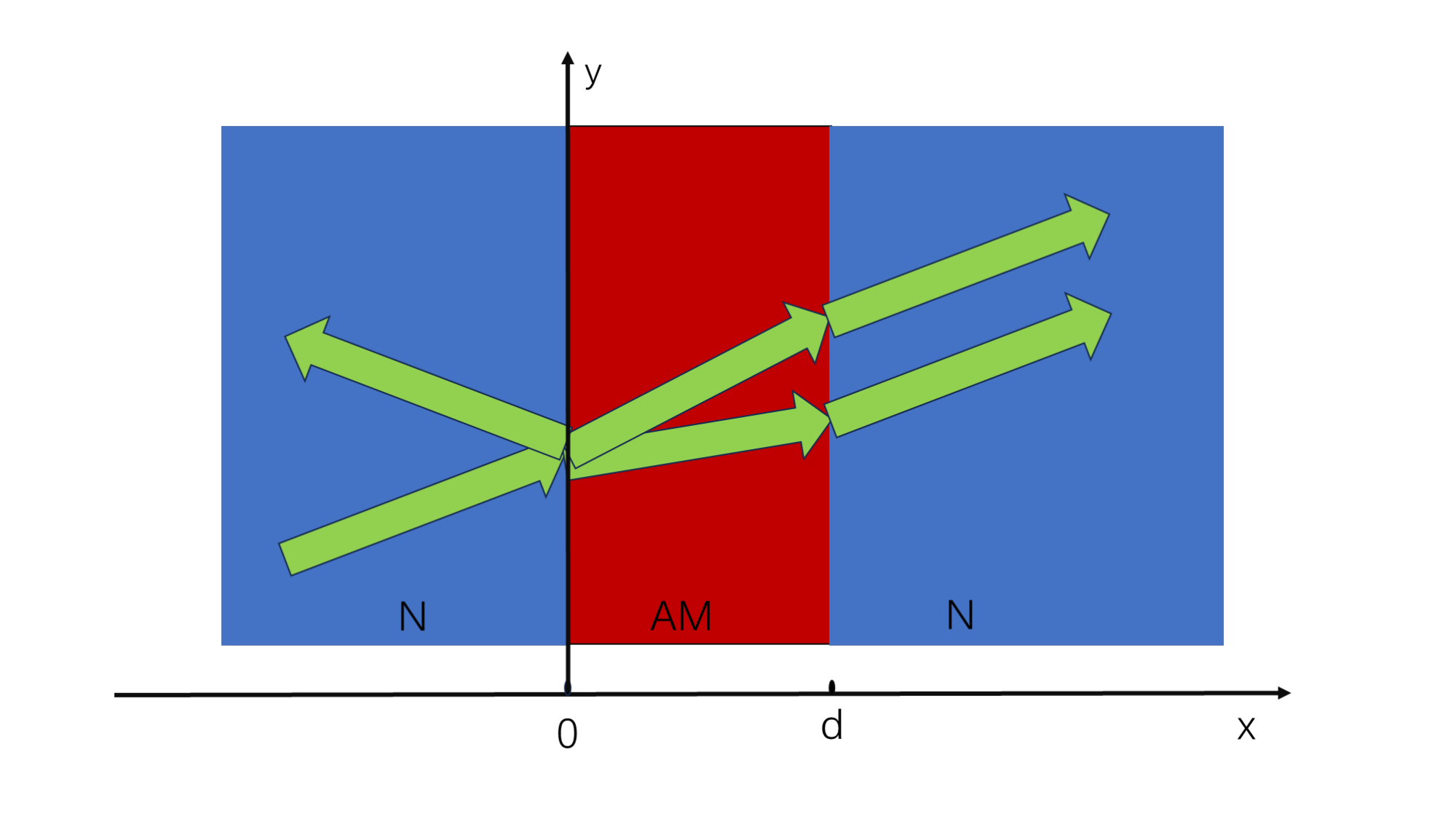}
	\caption{Schematic diagram of two dimensional heterojunction consisted of Normal metal/p-wave altermagnet/Normal metal structure.  }
	\label{fig1}
\end{figure}

We consider a heterostructure geometry in the $(x, y)$ plane shown in Fig. \ref{fig1}. The system consists of three regions: the left normal metal (N) region $(x<0)$ occupying the negative $x$ -axis, the intermediate p-wave altermagnet (AM) region $(0<x<d)$ of finite width, and the right normal metal (N) region $(x>d)$ occupying the positive $x$ -axis. The two interfaces are located at $x=0$ and $x=d$, respectively.

The system is assumed to be translationally invariant in the $y$ direction, and the transport is in the ballistic regime. The normal metal regions can be approximated as a two-dimensional semi-infinite electron gas, with low-energy excitations described by a parabolic band $\mathbf{H}_{I,III}(k) = t_0 k^2 \sigma_0$, with $t_0$ is the transfer integral and $\sigma_0$ is the $2\times2$ identity matrix. $p$-wave altermagnet is a pure electronic nonrelativistic spin-splitting. It is generated in chiral spin helix phase in which spin form helical structure\cite{REF23}. Elastic scattering at the interface allows for conservation of the total energy and of the momentum parallel to the interface.
The low-energy effective Hamiltonian of the p-wave altermagnet region can be written as

\begin{equation}
\label{eq-Hamiltonian}
\mathbf{H}(\mathbf{k}) = t_0 k^2 \sigma_0 + \mathbf{h}(\mathbf{k}) \cdot \boldsymbol{\sigma}
\end{equation}

Here $\mathbf{h}(\mathbf{k}) = (p k_x, \lambda k_y, m_z)$. In which $p$  is the strength of spin-splitting applied to lift the degeneracy of bands on the spin-nodal plane. A relativistic spin-orbit coupling $\lambda$ is introduced to allow for breaking inversion-symmetry. $\boldsymbol{\sigma}(\sigma_x,\sigma_y,\sigma_z)$ are Pauli matrix for spin.

This Hamiltonian satisfies the symmetry conditions of a $p$-wave alternating magnet, namely that time-reversal symmetry is preserved while spatial inversion symmetry is broken, which is manifested as a horizontal shift of the spin-split bands in momentum space along a specific direction\cite{REF24}.

In normal metallic region$(x<0)$, spin degeneracy is observed. Energy dispersion is isotropic parabola $\varepsilon _q =t_0 q^2 $.

When electron incident from region I with angle $\theta_i$, the components of momentum along and perpendicular to the horizon are $q_x = q \cos\theta_i$, $q_y = q \sin\theta_i$, respectively.
 The wave function in this region can be written as \cite{zhang2005spinselected}

\[
\Psi_I = \left[ \frac{1}{\sqrt{2}} \begin{pmatrix} 1 \\ 1 \end{pmatrix} e^{iq_x x} + \begin{pmatrix} A_1 \\ A_2 \end{pmatrix} e^{-iq_x x} \right] e^{iq_y y}
\]

The first term represents incident spinor base. The second term corresponds to reflective wavefunction. $A_1$,$A_2$ are unknown variables.

The transmission wave function in region III $(x>d)$ has the form
\[
\Psi_I = \left[  \begin{pmatrix} G_1 \\ G_2 \end{pmatrix} e^{iq_x x} \right] e^{iq_y y}
\]

In $p$-wave altermagnet region $(0<x<d)$, spin degeneracy is lifted. Due to the presence of spin-orbit coupling and $p$-wave spin splitting, the spin-up and spin-down states cannot be completely separated.
Hamiltonian in Eq.\eqref{eq-Hamiltonian} can be written in matrix form,

\begin{equation}
\label{eq-H_MATRIX}
H(k) =
\begin{pmatrix}
t_0k^2 + m_z & pk_x - i\lambda k_y \\
pk_x + i\lambda k_y & t_0k^2 - m_z
\end{pmatrix}
\end{equation}

Solving the equation $H\Psi=E\Psi$, we obtain electron energy,
\begin{equation}
\label{eq-Energy}
E=t_0 k^2 \pm \sqrt{m_z ^2+(p k_x)^2+(\lambda k_y)^2}.
\end{equation} .

Here $\prime+\prime$, $\prime-\prime$ represent different mode. Owing to translation invariance in $y$ direction, $y$ component of wave vector is conserved when crossing the interface, $k_y=q_y$. For specific indient energy and transverse wave vector, longitudinal wave vector of $p$-wave altermagnet is determined by energy conservation and the mode. They can be solved from Eq.\eqref{eq-Energy},

\begin{equation}
\label{eq-k square}
k_{\pm}^{2} = \frac{(2Et + p^{2}) \mp \sqrt{(2Et + p^{2})^{2} - 4t^{2}[E^{2} - m_{z}^{2} + (p^{2} - \lambda^{2})k_{y}^{2}]}}{2t^{2}}
\end{equation}

The longitudinal component is obtained as
\[
k_{\pm ,x}=\sqrt{k_{\pm}^2-k_y ^2}.
\]

When the expression under the square root becomes negative, the corresponding longitudinal wave vector becomes imaginary, indicating that electrons in this spin channel exist as evanescent waves in the alternating magnet and cannot propagate. Physically, this condition corresponds to a spin-selective total reflection effect.

Due to symmetry along $y$ direction, the $y$-component of momentum is conserved before and after scattering. The following relations are naturally arrived

\begin{equation}
\label{eq-ky}
q_y = q \sin \theta_i = k_- \sin \theta_- = k_+ \sin \theta_+ = k_y
\end{equation}

As $k_+ \neq k_-$, the two branches in AM refract with different angles as shown in Fig. \ref{fig1}. The refractive angles of the two branches are obtained from Eq. \eqref{eq-ky},

\[
\sin \theta_+ = \frac{k_y}{k_+}, \sin \theta_- = \frac{k_y}{k_-}.
\]

The incident particle occurs birefringence in the p-wave magnet region. $'+'$ branch electron and $'-'$ electron have different refraction angles, which means different propagation directions. When they reach the second interface and enter the normal metal again, transmission waves propagate in the same direction as the incident wave with certain transversal shift depending on the incident angle and AM length.

In $p$-wave altermagnet, electron wave function can be formally written as superposition of two propagation modes. Each mode has right-moving part and left-moving part due to scattering at interface.

\begin{equation}
\label{eq-PsiII}
\begin{split}
\Psi_{II} = \Bigg\{ & B \begin{pmatrix} 1 \\ z_1 \end{pmatrix} e^{ik_x^+ x} + C \begin{pmatrix} 1 \\ z_2 \end{pmatrix} e^{-ik_x^+ x} \\
& + D \begin{pmatrix} 1 \\ z_3 \end{pmatrix} e^{ik_x^- x} + F \begin{pmatrix} 1 \\ z_4 \end{pmatrix} e^{-ik_x^- x} \Bigg\} e^{ik_y y}
\end{split}
\end{equation}

Where $B$,$C$,$D$,$F$ are unknown coefficients. Some intermediate variables $z_1$, $z_2$, $z_3$, $z_4$ are introduced to shorten the whole expression.

\begin{equation}
\begin{split}
  &z_1 = \frac{\sqrt{m_z^2 + (p k_{+,x})^2 + (\lambda k_y)^2} - m_z}{p k_{+,x} - i\lambda k_y},\\
  &z_2 = \frac{\sqrt{m_z^2 + (p k_{+,x})^2 + (\lambda k_y)^2} - m_z}{-p k_{+,x} - i\lambda k_y},\\
  &z_3 = \frac{\sqrt{m_z^2 + (p k_{-,x})^2 + (\lambda k_y)^2} + m_z}{-p k_{-,x} + i\lambda k_y},\\
  &z_4 = \frac{\sqrt{m_z^2 + (p k_{-,x})^2 + (\lambda k_y)^2} + m_z}{p k_{-,x} + i\lambda k_y}.
\end{split}
\end{equation}

Longitudinal velocity of each mode in $p$-wave altermagnet is

\begin{equation}
\label{eq-velocity2}
v_{\pm,x} = \frac{1}{\hbar} \frac{\partial H}{\partial k_x}
= \frac{1}{\hbar}
\begin{pmatrix}
2 t_0 k_{\pm,x} & p \\
p & 2 t_0 k_{\pm,x}
\end{pmatrix}
\end{equation}

At the interface of $x=0$ and $x=d$, wave function and their derivatives are continuous. This continuous condition arises from Schr\"{o}dinger equation integral around the interface. In ballistic transport and nonbarrier condition, the continuous condition is expressed as

\begin{equation}
\label{eq-continuity}
\begin{split}
&\Psi_I(x = 0^-) = \Psi_{II}(x = 0^+), \\
&\Psi_{II}(x = d^-) = \Psi_{III}(x = d^+), \\
&[-t\sigma_0 \Psi'(0^+) + t_0 \sigma_0 \Psi'(0^-)] - ip\sigma_x[\Psi(0^+) - \Psi(0^-)] = 0, \\
&[-t_0 \sigma_0 \Psi'(d^+) + t\sigma_0 \Psi'(d^-)] - ip\sigma_x[\Psi(d^+) - \Psi(d^-)] = 0.
\end{split}
\end{equation}

Substituting wave functions $\Psi_I$, $\Psi_{II}$, and $\Psi_{III}$  into Eq.\eqref{eq-continuity}, we obtain the following combined linear equations,

\begin{equation}
\label{eq-Linear}
\begin{split}
&-A_1 + B + C + D + F = \frac{1}{\sqrt{2}} \\
&-A_2 + z_1 B + z_2 C + z_3 D + z_4 F = \frac{1}{\sqrt{2}} \\
&e^{ik_{+,x}d} B + e^{-ik_{+,x}d} C + e^{ik_{-,x}d} D + e^{-ik_{-,x}d} F - e^{iq_x d} G_1 = 0 \\
&z_1 e^{ik_{+,x}d} B + z_2 e^{-ik_{+,x}d} C + z_3 e^{ik_{-,x}d} D + z_4 e^{-ik_{-,x}d} F - e^{iq_x d} G_2 = 0 \\
&\alpha_1  B + \alpha_2 C + \alpha_3 D  + \alpha_4 F + \alpha_0 A_1 = \frac{\alpha_0}{\sqrt{2}}  \\
&\beta_1 B + \beta_2 C + \beta_3 D  + \beta_4 F + \alpha_0 A_2 =  \frac{\alpha_0}{\sqrt{2}}  \\
&\alpha_1 e^{ik_{+,x}d} B + \alpha_2 e^{-ik_{+,x}d} C + \alpha_3 e^{ik_{-,x}d} D \\
&\quad  + \alpha_4 e^{-ik_{-,x}d} F - \alpha_0 e^{iq_x d} G_1 = 0 \\
&\beta_1 e^{ik_{+,x}d} B + \beta_2 e^{-ik_{+,x}d} C + \beta_3 e^{ik_{-,x}d} D \\
&\quad  + \beta_4 e^{-ik_{-,x}d} F - \alpha_0 e^{iq_x d} G_2 = 0
\end{split}
\end{equation}

In this equation, $X=(A_1,A_2,B,C,D,F,G_1,G_2)^T $ are the unknown variables we are looking for. $\alpha_{i}$ and $\beta_{i}(i=1,2,3,4,0)$  are used to shorten the length of the formula, with $\alpha_0 = t_0 q_x$, $\alpha_1 = (t_0 k_{+,x} + pz_1)$, $\alpha_2 = (-t_0 k_{+,x} + pz_2)$, $\alpha_3 = (t_0 k_{-,x} + pz_3)$, $\alpha_4 = (-t_0 k_{-,x} + pz_4)$, $\beta_1 = (t_0 k_{+,x} z_1 + p)$, $\beta_2 = (-t_0 k_{+,x} z_2 + p)$, $\beta_3 = (t_0 k_{-,x} z_3 + p)$, $\beta_4 = (-t_0 k_{-,x} z_4 + p)$.
These equations can be transformed into matrix form as $MX=W$ , where $W$ is the column of constant terms and $M$ is the coefficient matrix, respectively. The full content of the matrix and arrays are

{\scriptsize
\begin{equation}
\label{eq-M matrix}
\begin{gathered}
M = \\
\begin{pmatrix}
-1 & 0 & 1 & 1 & 1 & 1 & 0 & 0 \\
0 & -1 & z_1 & z_2 & z_3 & z_4 & 0 & 0 \\
0 & 0 & e^{ik_{+,x} d} & e^{-ik_{+,x}d} & e^{ik_{-,x} d} & e^{-ik_{-,x}d} & -e^{iq_x d} & 0 \\
0 & 0 & z_1 e^{ik_{+,x} d} & z_2 e^{-ik_{+,x} d} & z_3 e^{ik_{-,x} d} & z_4 e^{-ik_{-,x} d} & 0 & -e^{iq_x d} \\
\alpha_0 & 0 & \alpha_1 & \alpha_2 & \alpha_3 & \alpha_4 & 0 & 0 \\
0 & \alpha_0 & \beta_1  & \beta_2  & \beta_3  & \beta_4  & 0 & 0 \\
0 & 0 & \alpha_1 e^{ik_{+,x}d} & \alpha_2 e^{-ik_{+,x}d} & \alpha_3 e^{ik_{-,x}d} & \alpha_4 e^{-ik_{-,x}d} & -\alpha_0 e^{iq_x d} & 0\\
0 & 0 & \beta_1 e^{ik_{+,x}d} & \beta_2 e^{-ik_{+,x}d} & \beta_3 e^{ik_{-,x}d} & \beta_4 e^{-ik_{-,x}d} & 0 & \alpha_0 e^{iq_x d}
\end{pmatrix}
\end{gathered}
\end{equation}
}

\[
W=(\frac{1}{\sqrt{2}},\frac{1}{\sqrt{2}},0,0,\frac{q_x t_0}{\sqrt{2}},\frac{q_x t_0}{\sqrt{2}},0,0)^T.
\]

All the variables are solved by numerically calculating  $X=M^{-1} W$. The reflection and transmission property of heterojunction can be analyzed from these variables.

In order to discuss spin-resolved transport property, we introduce flux density. Velocity in segment I and III are obtained,
\[
v_{ix} =
\begin{pmatrix}
2t_0q_x & 0 \\
0 & 2t_0q_x
\end{pmatrix}
\]
\[
v_{tx} =
\begin{pmatrix}
2t_0q_x & 0 \\
0 & 2t_0q_x
\end{pmatrix}
\]

The incident and reflection flux density are defined as
\[
J_i = \langle \Psi_i|v_{ix}|\Psi_i\rangle =2t_0 q_x,
\]
\[
J_r = \langle \Psi_r|-v_{ix}|\Psi_r\rangle =-2t_0 q_x,
\]
The minus sign represents refractive flux to the left. It can be omitted if we focus on the flux magnitude only.
\[
J_{r\uparrow} = 2t_0 q_x |A_1|^2,
\]
\[
J_{r\downarrow} = 2t_0 q_x |A_2|^2,
\]

Reflection probability is defined as
\[
R_{\uparrow} = \frac{J_{r\uparrow}}{J_i} =|A_1|^2,
\]
\[
R_{\downarrow} = \frac{J_{r\downarrow}}{J_i} =|A_2|^2,
\]

In the same way, transmission flux density and transmission probability are calculated by the following formula
\[
J_{t\uparrow} = \langle\Psi_{t\uparrow}|v_{tx}|\Psi_{t\uparrow}\rangle,
\]
\[
J_{t\downarrow} = \langle\Psi_{t\downarrow}|v_{tx}|\Psi_{t\downarrow}\rangle,
\]
\[
T_{\uparrow} = \frac{J_{t\uparrow}}{J_i} =|G_1|^2,
\]
\[
R_{\downarrow} = \frac{J_{t\downarrow}}{J_i} =|G_2|^2,
\]

From flux density conservation, we obtain $R_\uparrow +R_\downarrow+T_\uparrow +T_\downarrow =1$. This is a rule to check the correctness of the calculation. Polarization of transmission is the key parameter describing spin-selective transmission. It is defined as ratio of difference of spin up probability with spin down probability and total transmission probability.
\[
P_J = \frac{J_{\uparrow}-J_{\downarrow}}{J_{\uparrow}+J_{\downarrow}}
\]
\[
P_T = \frac{T_{\uparrow}-T_{\downarrow}}{T_{\uparrow}+T_{\downarrow}}
\]

Naturally, $P_J$ is equal to $P_T$. The value of $P_T$ is in the range of $-1.0$ to $1.0$. $P_T =0$ means unpolarized. $P_T = \pm 1$ means transmission is fully polarized.

In normal metal, $P_T$ is always zero due to spin degeneracy.  In an N/AM/N junction formed by a $p$-wave altermagnet, the alternating magnet produces opposite momentum offsets to electrons of different spins. Consequently, after the incident electrons propagating through the entire structure, the transmission amplitudes and accumulated phases of the two spin components become different. This results in a nonzero transmission polarization. In physical essence, this behavior is analogous to the spin birefringence effect in graphene. However, its driving mechanism originates from non-relativistic lattice symmetry breaking rather than spin-orbit coupling \cite{REF25}.

The above theoretical framework provides a complete quantum-transport description for the birefringence effects in $p$-wave altermagnets. By tuning parameters such as the altermagnet junction length $d$ , the staggered splitting strength $p$ , or the incident angle $\theta_i$ , the reflection and transmission probabilities as well as the transmission polarization exhibit rich physical behaviors. The underlying symmetry analysis and tunability will be the focus of the next section.

\subsection{Birefringence at N/AM Interface}
When the system is consisted of normal metallic region $(x<0)$ and semi-infinite $p$-wave altermagnet $(x>0)$, with only one interface at $x=0$, the transport problem seems simpler in form. However, the physical essence tells the central mechanism of spin birefringence in altermagnet. In this structure, electron is incident from normal metallic part to the interface. Reflection and refraction happen at the interface. Since the second interface doesn't exist, transmitted waves can propagate to the infinite without reflection.

Different from the above derivative of heterojunction, wave function $\Psi_{III}$ disappears. The left moving terms in $\Psi_{II}$ also disappear because no reflection in the second part.

\[
\Psi_{II} = \left[ B \begin{pmatrix} 1 \\ z_1 \end{pmatrix} e^{ik_{+,x} x} +D\begin{pmatrix} 1 \\ z_3 \end{pmatrix} e^{ik_{-,x} x }  \right] e^{ik_y y}
\]

However, wave functions and their derivatives are continuous at the interface. Matrix form equation becomes

\begin{equation}
\begin{pmatrix}
-1 & 0 & 1 & 1 \\
0 & -1 & z_1 & z_3 \\
\alpha_0 & 0 & \alpha_1 & \alpha_3 \\
0 & \alpha_0 & \beta_1& \beta_3
\end{pmatrix}
\begin{pmatrix}
A_1 \\
A_2 \\
B \\
D
\end{pmatrix}
=
\begin{pmatrix}
1 / \sqrt{2} \\
1 / \sqrt{2} \\
\alpha_0 / \sqrt{2} \\
\alpha_0 / \sqrt{2}
\end{pmatrix}.
\end{equation}

Although the number of unknown variables becomes less, scattering at the interface is more complex, since there are both $'+'$ mode and $'-'$ mode in the $p$-wave altermagnet part. Obviously, velocities of the two modes are different.

\[
v_{+,x} =\frac{1}{\hbar}
\begin{pmatrix}
2t_0k_{+,x} & p \\
p & 2t_0 k_{+,x}
\end{pmatrix}
\]
\[
v_{-,x} =\frac{1}{\hbar}
\begin{pmatrix}
2t_0 k_{-,x} & p \\
p & 2t_0 k_{-,x}
\end{pmatrix}
\]

Flux densities of the modes are also different.
\[
J_{+} = \langle\Psi_{+}|v_{+,x}|\Psi_{+}\rangle,
\]
\[
J_{-} = \langle\Psi_{-}|v_{-,x}|\Psi_{-}\rangle,
\]

From these flux densities we can calculate transmission probability of the two modes.
\[
T_{+}=\frac{J_{+}}{J_i}
\]
\[
T_{-}=\frac{J_{-}}{J_i}
\]

Scattering at the interface also conserves flux density, $R_{\uparrow} +R_{\downarrow} +T_{+}+T_{-}=1$.
With these theoretical preparations, we are ready to do numerical calculations and analyze numerical results.

\section {Numerical Calculation and Discussions }

\subsection{ Birefringence at N/AM interface}
\begin{figure}
	\includegraphics[width=0.9\columnwidth]{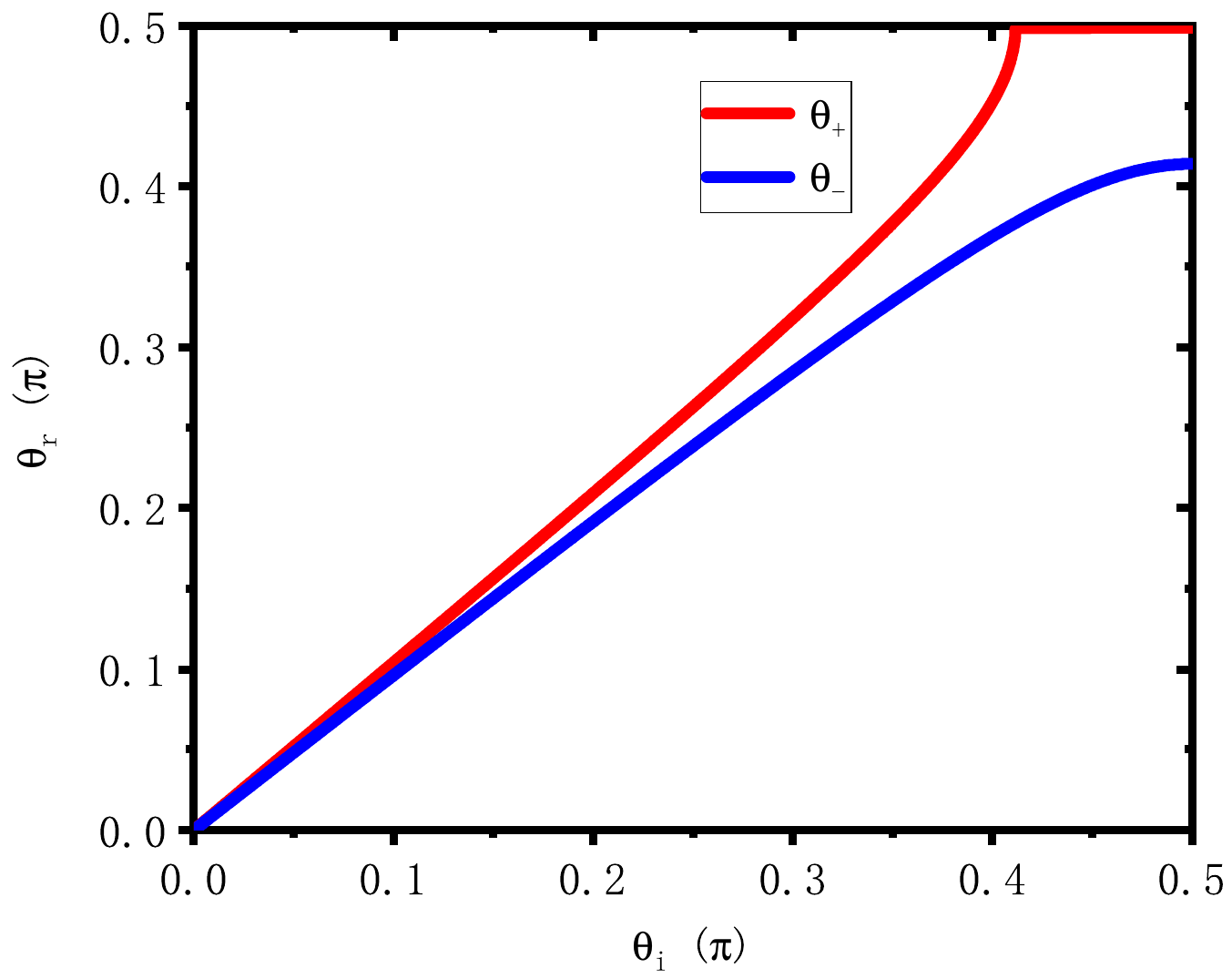}
	\caption{Refraction angles vs incident electron with parameters $p=0.1$, $\lambda=0.1$, $m_z =0.2$, $E=3.0$.}
	\label{fig02}
\end{figure}

Electron with energy E incident from normal metal to the $p$-wave AM. Reflection and refraction occur at the interface. $'+'$ branch and $\prime-\prime$ branch separate in AM material. Fig. \ref{fig02} plots refractive angles change with incident angle. Both $\theta_+$ and $\theta_-$ increase with $\theta_i$. The relation of $k_+$, $k_-$ and $q$ results in $\theta_+ >\theta_i$, $\theta_- <\theta_i$. Birefringence happens when electrons incident from normal metal region and propagate to AM region. When $\theta_i$ reaches a specific angle, $\theta_+$ reaches $\pi /2$ while $\theta_-$ saturates to a constant value. This reflects total reflection for $'+'$ branch. The specific incident angle is called critical angle, denoted as $\theta_c$.

\begin{figure}
	\includegraphics[width=0.9\columnwidth]{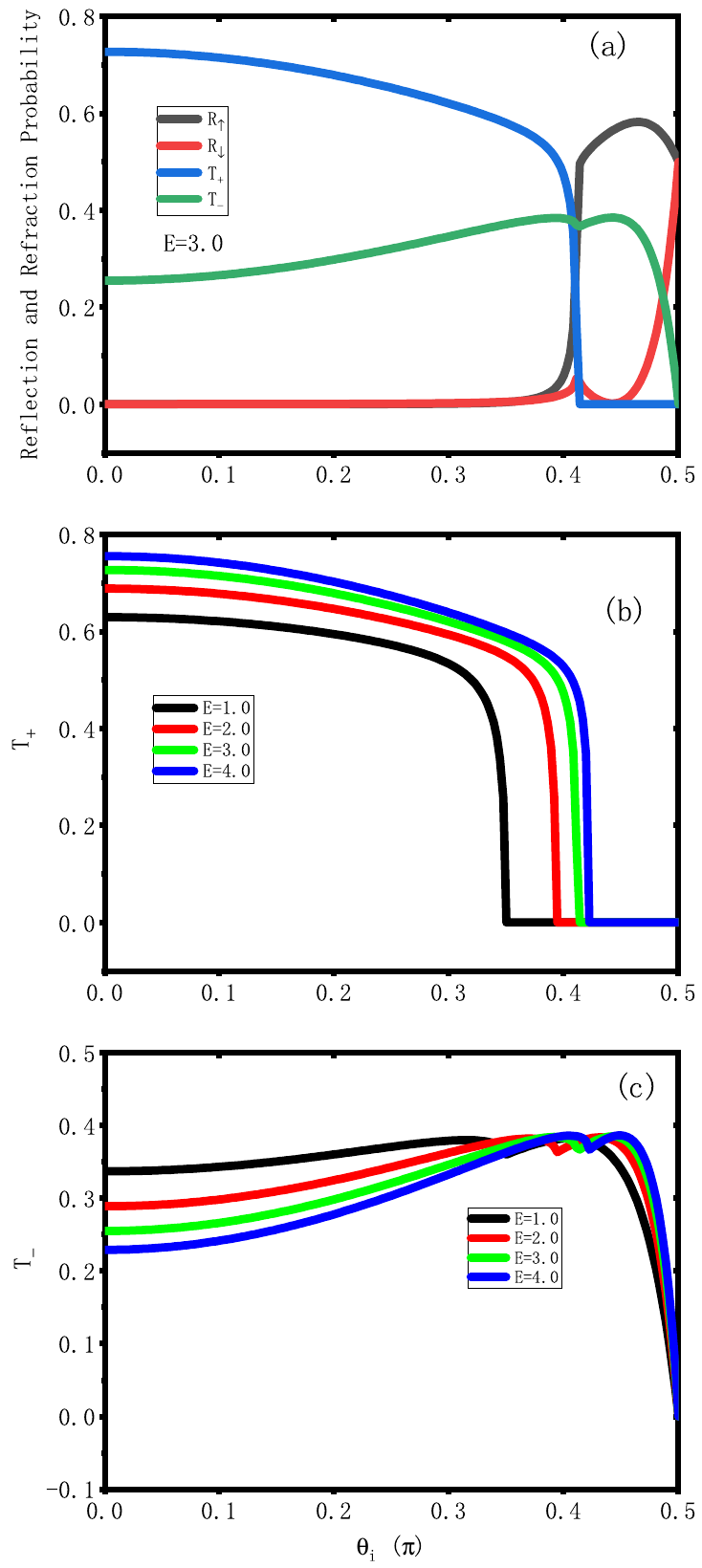}
	\caption{Scattering at the interface vs incident angle. (a) Reflection and refraction probability change with incident angle for electron energy $E=3.0$. (b), (c) Transmission probability for $'+'$ and $\prime-\prime$ branches change with incident angle. Parameters are  $p=0.1$, $\lambda=0.1$, $m_z =0.2$ }
	\label{fig03}
\end{figure}

Similar to the reflection and refraction of light at an interface, when the incident angle is small, the reflection coefficient is very small while the transmission coefficient is relatively large, as shown in Fig.\ref{fig03}. When the incident angle approaches approximately $0.4\pi$, the reflection coefficient for the spin-up component suddenly increases rapidly, while the transmission coefficient for the $'+'$ branch drops sharply to zero. This is because the $'+'$ branch reaches the condition for total reflection. This also indicates that the transmitted $'+'$ branch contains a significant portion of spin-up components. In contrast, the transmitted $\prime-\prime$ branch contains more spin-down components, as the transmittance of the refracted $\prime-\prime$ branch exhibits a trend exactly opposite to that of the spin-down reflectance. The distinct turning point near $0.4\pi$ reflects the spin-flip coupling between spin-up and spin-down states induced by spin-orbit coupling.
Panels (b) and (c) show how the refractive coefficients of the two branches vary with the incident angle under different incident energies. The larger the incident energy, the larger the critical angle for total reflection of the $'+'$ branch. A preliminary conclusion is drawn that the critical angle $\theta_c$ is determined by incident energy. When $\theta_i < \theta_c$, a larger incident energy corresponds to a larger refractive coefficient for the $'+'$ branch, but a smaller refractive coefficient for the $\prime-\prime$ branch. At the critical angle $\theta_c$, the refractive coefficient of the $'+'$ branch drops sharply to zero, while that of the $\prime-\prime$ branch only decreases slightly and does not drop rapidly to zero until at larger incident angles.

\begin{figure}
	\includegraphics[width=0.9\columnwidth]{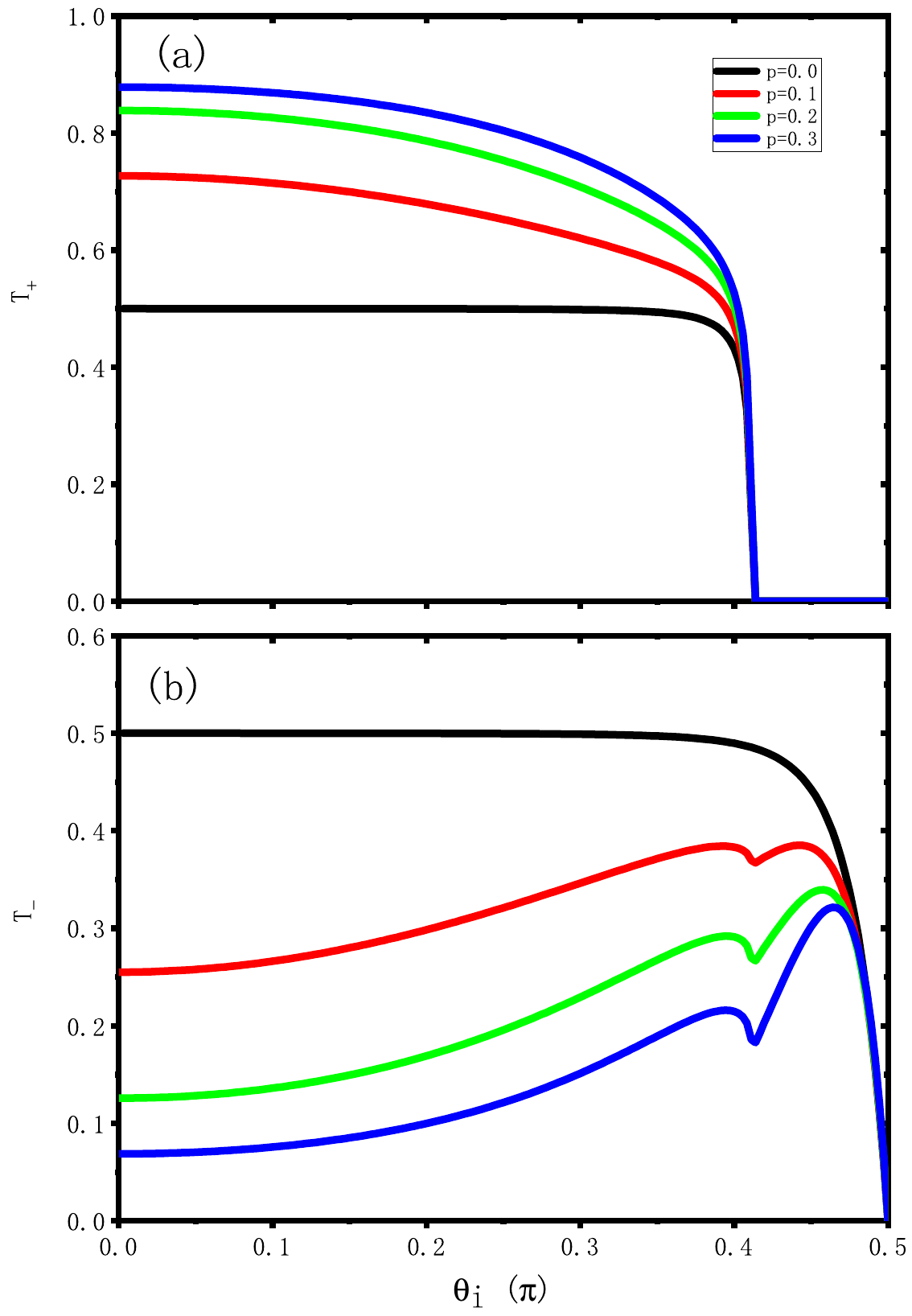}
	\caption{Refraction coefficient changes with incident angle with different spin-splitting strength p in altermagnet. Other parameters are set $\lambda=0.1$, $m_z =0.2$, $E=3.0$.}
	\label{fig04}
\end{figure}

Fig. \ref{fig04} (a) and (b) show refraction coefficients of $'+'$ and $\prime-\prime$ branches when changing spin-splitting strength. Indeed, the refractive coefficients are significantly affected by spin-splitting strength $p$. The larger the value of $p$, the larger the refractive coefficient of the $'+'$ branch, but the correspondingly smaller the refractive coefficient of the $\prime-\prime$ branch. However, $T_+$ curves drop to zero at the same angle, no matter how strong the spin-splitting is in the AM. This implies spin-splitting strength $p$ does not affect the critical angle for total reflection of the $'+'$ branch.

\begin{figure}
	\includegraphics[width=0.9\columnwidth]{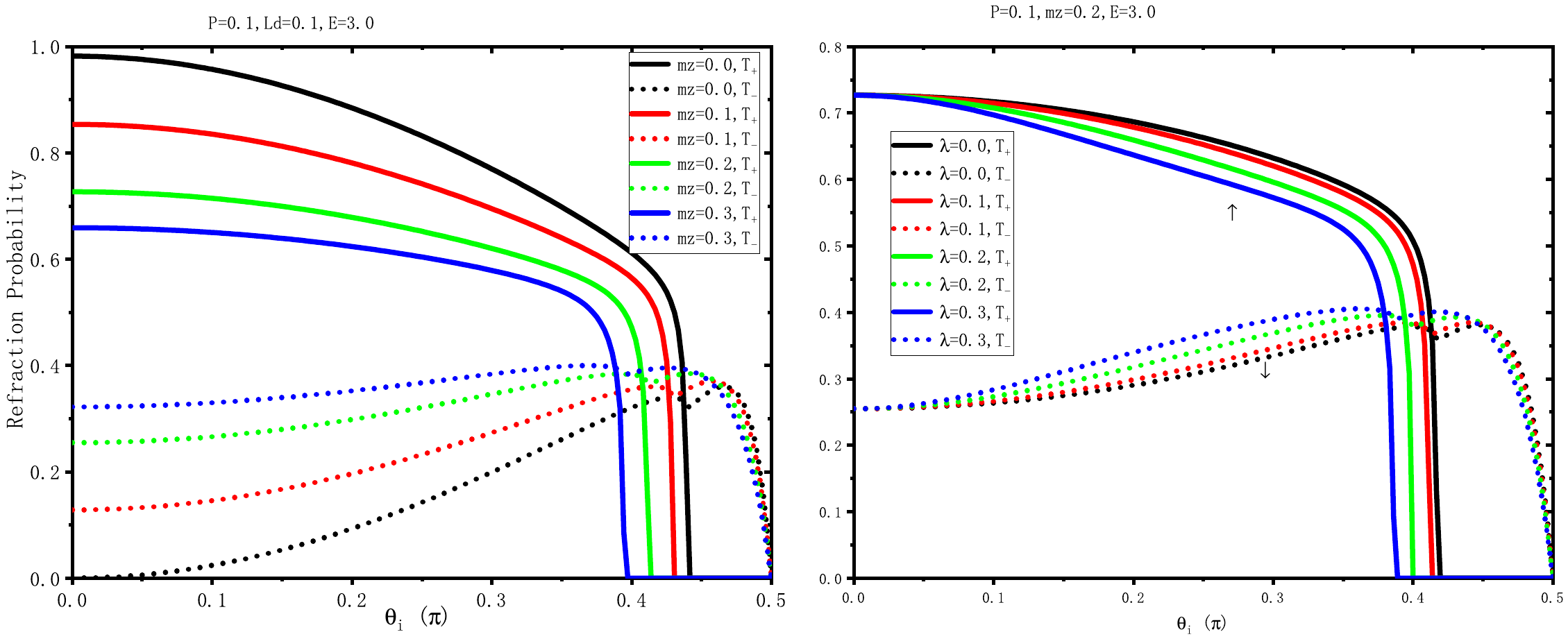}
	\caption{Refraction coefficients as a function of incident angle when varying exchange field and spin-orbit coupling. Solid lines denote transmission of $'+'$ branch, while dashed lines signify $\prime-\prime$ branch.  Other parameters for left panel $p=0.1$,$\lambda=0.1$, $E=3.0$.For right panel $p=0.1$, $m_z =0.2$, $E=3.0$.}
	\label{fig05}
\end{figure}

 Exchange field $m_z$ and spin-orbit coupling $\lambda$ of the altermagnet also have a pronounced influence on the refractive coefficients of $'+'$ and $\prime-\prime$ branches. Fig. \ref{fig05} (a) is transmission changes with incident angle when varing strength of spin-orbit coupling $\lambda$. The refractive coefficients change accordingly, but the magnitude of the change is very small. In particular, when the incident angle is very small, the refractive coefficients remain almost unchanged. Only at larger incident angles do the refractive coefficients exhibit some variation with the spin-orbit coupling strength. Specifically, as the coupling strength increases, the refractive coefficient of the $'+'$ branch decreases slightly, while that of the $\prime-\prime$ branch increases slightly. The spin-orbit coupling strength causes a larger difference in the transmission properties of the two spin species in the altermagnet. Based on the above findings, it is revealed that the incident electron energy, the incident angle, the spin-splitting strength of the altermagnet, and the exchange field strength all have significant effects on the refractive coefficients of the $'+'$ and $\prime-\prime$ branches in the altermagnet, whereas the effect of the spin-orbit coupling is minor. The role of spin-orbit coupling is to generate spin-flip processes.

 When only $m_z$ is varied while all other parameters are kept the same, a smaller $m_z$ corresponds to a somewhat larger wave vector $k_+$, but $k_+$ is always smaller than the incident wave vector $q$. As a result, the critical angle for total reflection becomes larger. Consequently, the reflection coefficient is small while the refractive coefficient is large, as shown in Fig.\ref{fig05}(b). As $m_z$ is gradually increased, the value of $k_+$ gradually decreases, leading to a smaller refractive coefficient and making it easier to reach the critical angle for total reflection. The variation trend for the $\prime-\prime$ branch is opposite. As $m_z$ increases, the value of $k_-$ gradually increases, and so does the refractive coefficient. However, regardless of how the parameters change, it always holds that $k_+ < k_-$. Therefore, the refractive coefficient of the $'+'$ branch is always larger than that of the $\prime-\prime$ branch.

In the case of single-interface scattering, the two branches in the altermagnet have different refraction angles, but the spin states are not fully separated. Although the two branches transmit the interface with different probabilities, total transmission of the two branches are very close to 1.0 before the incident angle reaches critical angle $\theta_c$. The behavior is independent of incident energy, spin-orbit coupling strength as well as spin-splitting strength of altermagnet. It is similar to the characteristic of Klein tunneling. In the following, we investigate the case of double interfaces, i.e., transport through a heterojunction, to observe spin separation and polarization.

\subsection{Spin-Polarized Transmission of N/AM/N Heterostructure}
Unpolarized electrons from a normal metal region are incident to the N/AM interface at a certain angle. Reflection and refraction occur at the first interface, generating the $'+'$ and $'-'$ branches inside the altermagnet (AM). These branches then encounter the second interface, where reflection and refraction take place again. Upon exiting the AM and entering the normal metal, the electrons are split into spin-up and spin-down components. Although the two spin components have the same exit angle, they acquire a lateral spatial shift relative to each other. The lateral shift distance of the outgoing electron beam and the degree of spin polarization are influenced by the incident electron energy, the incident angle, the spin-splitting strength and the width of the altermagnet layer.
Firstly, let's take a look at how reflection and transmission probabilities change with AM length when incident angles are different.

\begin{figure}
	\includegraphics[width=0.9\columnwidth]{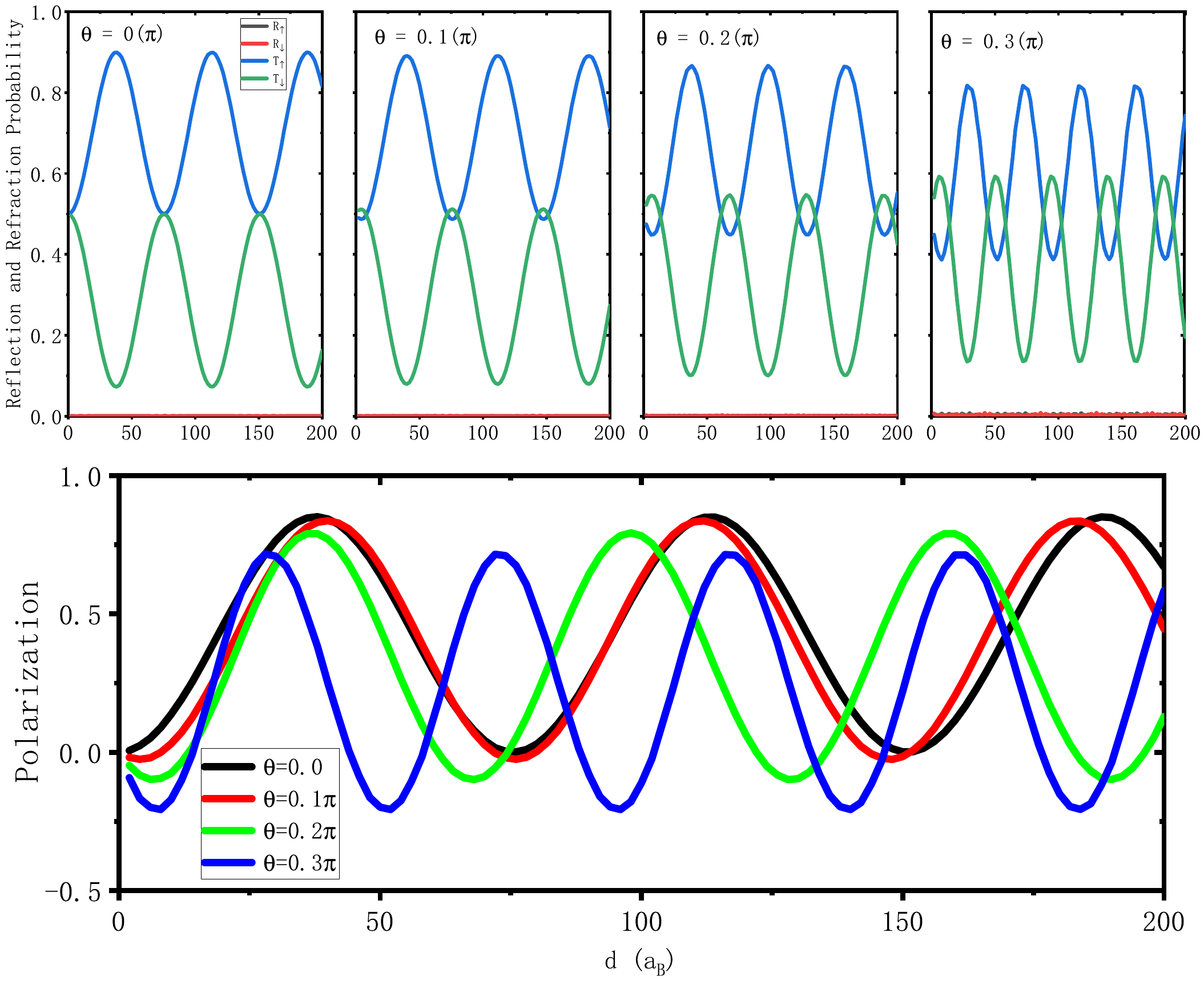}
	\caption{Transmission and polarization vary with incident d at different incident angles. Other parameters are fixed $p=0.1$,$\lambda=0.1$, $m_z =0.2$,$E=3.0$.}
	\label{fig06}
\end{figure}

The upper panels of Fig. \ref{fig06} show transmission and reflection probabilities change with AM length with different incident angles. In all cases, reflection probabilities are quite small, so we omit discussing reflections.
When incident angle is zero, transmission curves of spin-up and spin-down do not cross each other. This is because the transverse component of the wave vector vanishes, so the spin-orbit coupling does not take effect and no spin-flip occurs. According to discussions in the theoretical part, longitudinal components of wave vector for spin-up and spin-down states are not equal, i.e. $k_{+,x}<k_{-,x}$. Analogous to refractive index in optics, the relation of quasi refractive index of spin-up and spin-down electrons is $n_{+}<n_{-}$. Under otherwise identical conditions, the larger the refractive index, the larger the reflectance and the smaller the transmittance. The $'+'$ branch contains more spin-up components, while the $\prime-\prime$ branch contains more spin-down components. Therefore, the transmittance for spin-up is always greater than that for spin-down. When the incident angle is nonzero, spin-orbit coupling takes into effect, spin-flip occurs in the propagation. As a result, transmission curves for spin-up and spin-down cross each other. As the incident angle increases, transverse component of wave vector increases accordingly. The effect of spin-flip becomes obvious. Therefore, transmission curves cross further.
Another significant feature is periodicity of transmission vs AM length $d$ with any incident angle. As incident angle increases, periods of transmission curves decrease. Furthermore, transmission of spin-up becomes smaller while transmission of spin-down increases.
The lower panel shows polarization of transmission. When $\theta_i$ is zero, polarization is always positive. For finite $\theta_i$, spin polarization decreases and becomes negative with increasing angle, implying enhanced spin-flip.

\begin{figure}
	\includegraphics[width=0.9\columnwidth]{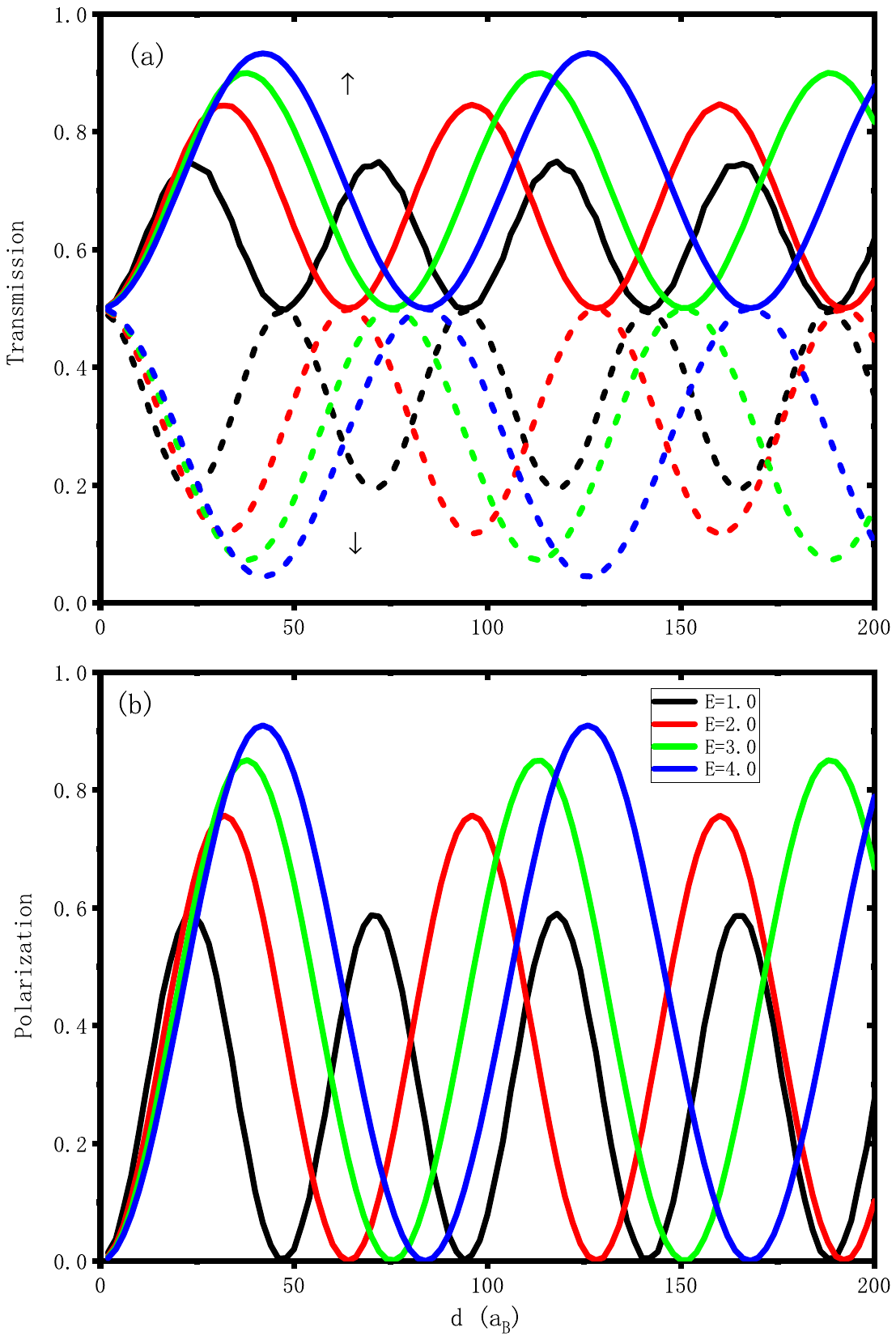}
	\caption{Transmission and polarization of electrons with different energies change with length of AM heterojunction. Parameters are set as $p=0.1$,$\lambda=0.1$, $m_z =0.2$,$\theta_i =0.0$}
	\label{fig07}
\end{figure}

Next, let's see transmission at vertical incident case with different incident energies, as shown in Fig. \ref{fig07}. Due to tiny reflection coefficient, only transmission is displayed. Panel (a) displays transmission varies with length of AM junction, with solid lines for spin-up and dashed lines for spin-down. Different colors represent cases of different incident energy. In the case of a specific incident energy, transmission coefficients of spin-up and spin-down electrons exhibit opposite trends when one increases, the other decreases. The transmittance for spin-up is always greater than or equal to $0.5$, while that for spin-down is always less than $0.5$. Moreover, the transmittance exhibits a periodic variation with the length of the altermagnet layer. The period depends on the incident energy. As the energy increases, the period of the transmittance variation becomes longer. The maximum transmittance for spin-up increases with increasing incident energy, while the minimum transmittance for spin-down decreases with increasing incident energy. Panel (b) shows the spin polarization of the transmitted electron beam. The spin polarization also varies periodically with the length. The larger the incident energy, the longer the period, and the higher the spin polarization that can be achieved. By choosing an appropriate length, one can obtain a transmitted electron beam with a high degree of spin polarization.

\begin{figure}
	\includegraphics[width=0.9\columnwidth]{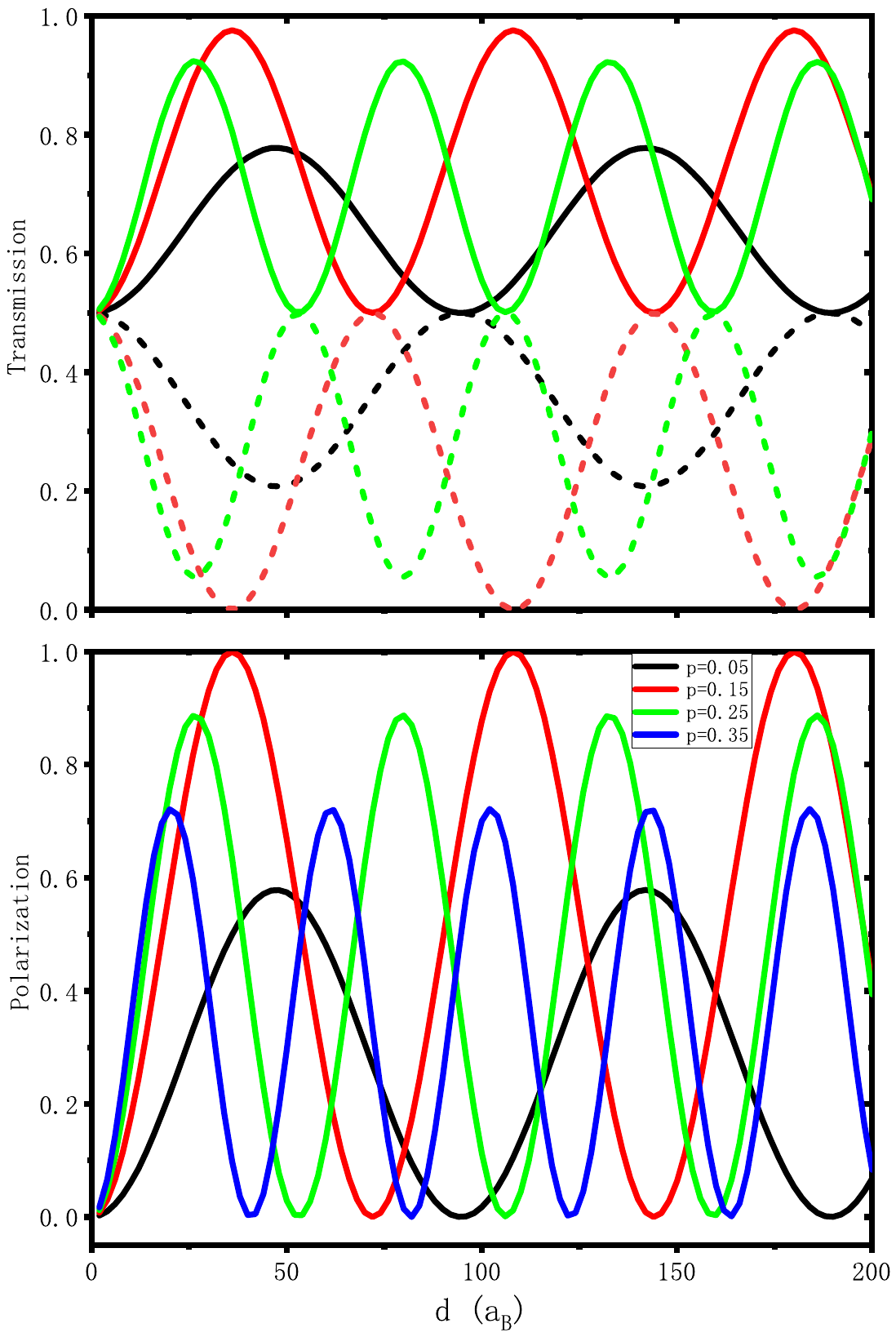}
	\caption{Transmission and polarization of electrons change with length of AM heterojunction with different spin-splitting strength. Parameters are set as $\lambda=0.1$, $m_z =0.2$,$E =4.0$,$\theta_i =0.0$}
	\label{fig08}
\end{figure}
With zero incident angle, both the transmittance and the spin polarization vary periodically with length $d$ of AM, as shown in Fig. \ref{fig08}. For a smaller value of spin-splitting strength $p$, the variations of the polarization and transmittance with $d$ are slower, and the oscillation period is larger. For a larger value of $p$, the variations become faster, and the period becomes smaller. The value of the spin polarization also changes accordingly. The polarization reaches a maximum value of $1.0$ when $p=0.15$. Under this condition, the transmittance for spin-down is zero, while that for spin-up is at its maximum, thus yielding the highest possible polarization, i.e., transmission of a single spin state.

\begin{figure}
	\includegraphics[width=0.9\columnwidth]{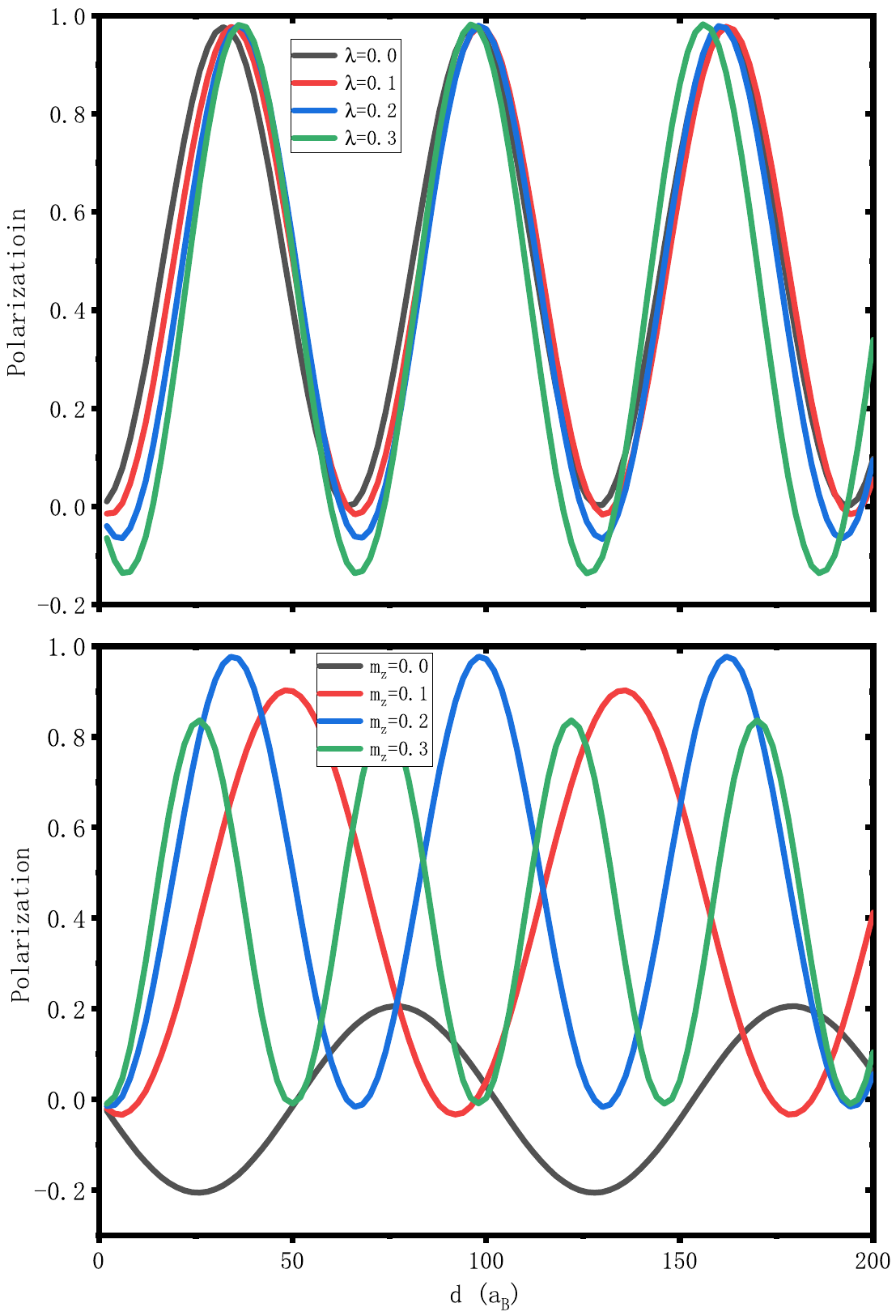}
	\caption{Spin polarization of transmission through AM heterojunction with various spin-orbit coupling and exchange filed. (a) $p=0.15$, $m_z =0.2$, $E =3.0$, $\theta_i =0.1\pi$. The black, red, blue and green curves represent the length-dependent polarization at spin-orbit coupling strengths of $0.0$, $0.1$, $0.2$, and $0.3$. (b)$p=0.15$, $\lambda=0.1$, $E =3.0$, $\theta_i =0.1\pi$. The curves in four colors correspond to the polarization as a function of length for exchange field strengths of $0.0$, $0.1$, $0.2$, and $0.3$, respectively.}
	\label{fig09}
\end{figure}

The upper panel of Fig. \ref{fig09} shows the spin polarization curves of the transmitted electron beam, which exhibit a periodic dependence on the length of the altermagnet (AM) junction. When only the spin-orbit coupling strength is varying, both the magnitude and the period of the spin polarization show no apparent change. In the presence of finite spin-orbit coupling, the transmission polarization curves periodically take negative values, which is a manifestation of spin-flip processes. At certain junction lengths, the spin polarization approaches unity, indicating single-spin electron beam obtained.
The lower panel displays spin polarization of transmission under the condition of varying exchange field $m_z$ only. Variation of $m_z$ has substantial effect on the polarization. When $m_z =0.0$, spin polarization is small and fluctuates around zero. This reflects the strong effect of spin-orbit coupling on spin-flip. When $m_z$ takes finite value, both the magnitude and period change apparently. In the case of $m_z = 0.2$, spin polarization approaches $1.0$. In experiment, single-spin transmission can be obtained by going through parameters in a certain range.

\begin{figure}
	\includegraphics[width=0.9\columnwidth]{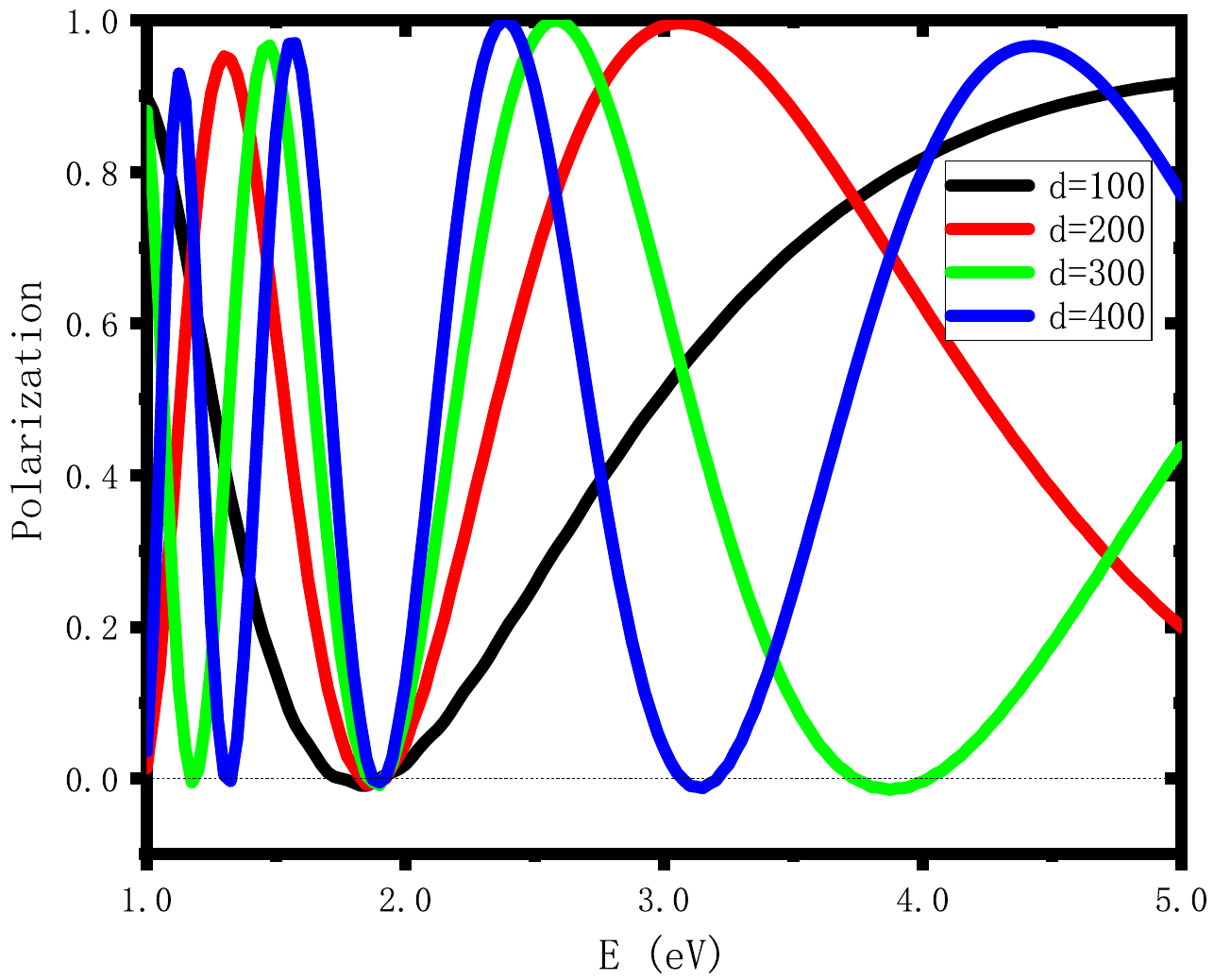}
	\caption{Spin polarization of transmission through AM heterojunction changes with energy. $p=0.2$, $\lambda=0.1$,$m_z =0.2$,$\theta_i =0.1\pi$.}
	\label{fig10}
\end{figure}

Fig. \ref{fig10} shows the curve of spin polarization of transmission varying with incident energy. Under certain conditions, the spin polarization curves fluctuate up and down, exhibiting oscillatory behavior. Polarization fluctuates fast in the low energy zone, but very slow when energy increases. For short heterojunctions, the polarization varies slowly with energy, whereas for long heterojunctions, the polarization varies frequently with energy. The polarization approaches unity at specific energy, revealing only spin-up transmitted. At some other energies, spin polarization drops to zero or even negative value, indicating bigger transmission of spin-down electron states.

\begin{figure}
	\includegraphics[width=1.0\columnwidth]{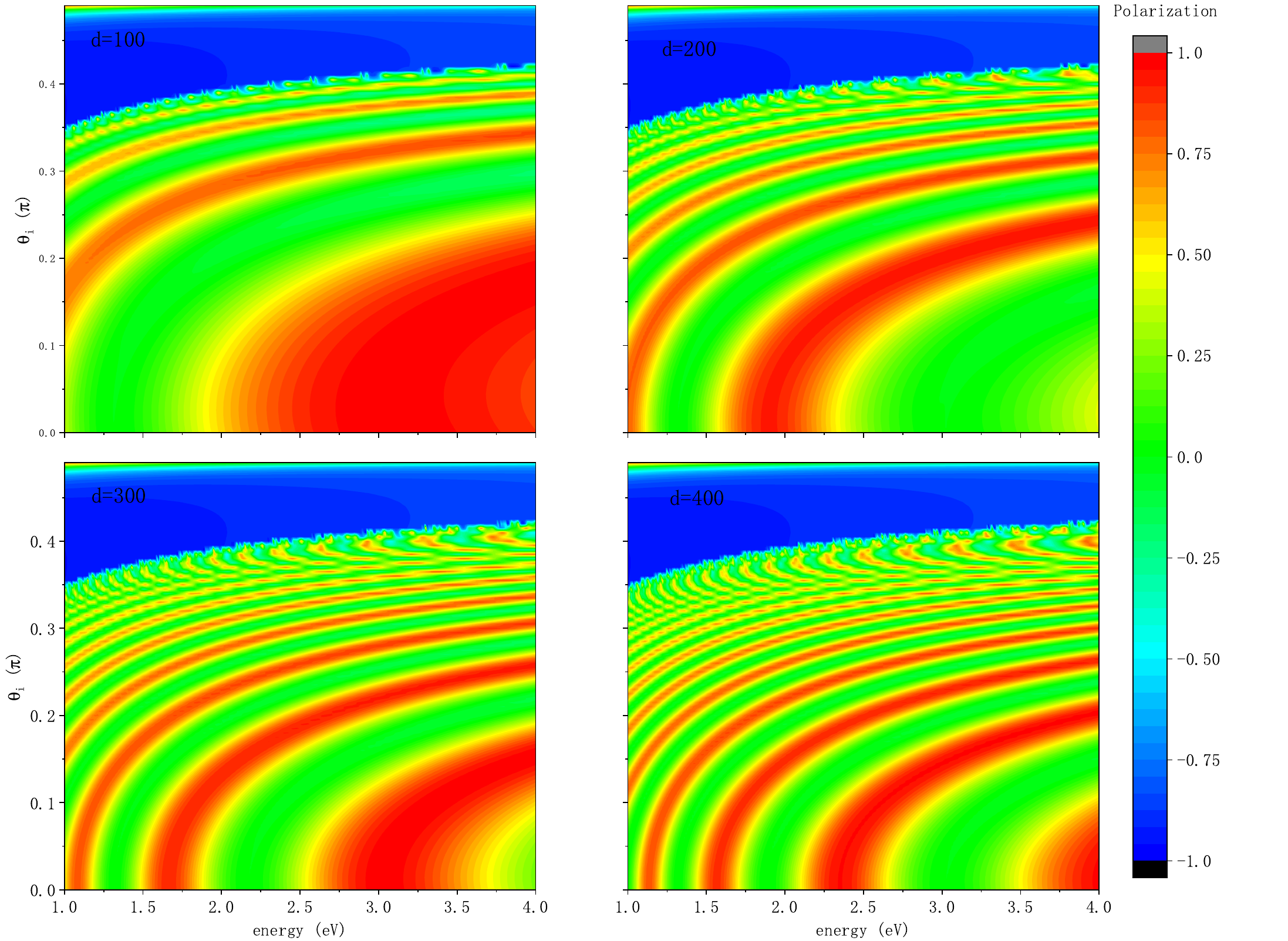}
	\caption{Transmission polarization changes with incident energy and incident angle. $p=0.15$, $\lambda=0.1$,$m_z =0.2$.}
	\label{fig11}
\end{figure}

The four panels in Fig. \ref{fig11} show the polarization corresponding to four different heterojunction lengths of $d=100$,  $d=200$,  $d=300$, and $d=400$, respectively. Up-left panel is the transmission polarization of heterojunction with length of $d=100$. It's conspicuous that transmission polarization changes at a slow pace when the length of junction is short. The blue part on the top represents that the polarization is $-1.0$. This is because the angle of incidence exceeds the critical angle, which results in only $\prime-\prime$ branch enters into altermagnet and transmits out. The red part in the bottom right corner indicates almost fully polarization. This corresponds to the condition of small incident angle and big energy. Green zone expresses unpolarized transmission. When the length of heterojunction increases, electrons incident into the altermagnet experience longer interaction distance of spin-splitting and spin-orbit coupling. This brings sufficient change of state and more fluctuations in polarization. The longer the junction, the more frequently the polarization changes. Polarization is positive in most of the range. It only takes small negative value when incident angle is close to the critical angle. This implies small possibility of spin-flip.

\section {Conclusion }
We investigate electron transport properties across N/AM interface and N/AM/N heterojunction. In the single interface case, when electron incident from normal metal to altermagnet, it is scattered and split into $'+'$ branch and $\prime-\prime$ branch transmitting into altermagnet zone with separate refractive angles. If incident angle is big enough, total internal reflection might happen to $'+'$ branch. This would not happen to $'-'$ branch. Critical angle of total reflection is dependent on incident conditions and properties of AM material.
When there exists a second interface, spin-up and spin-down electrons have same propagation directions with different transmission coefficients, depending on parameters such as incident angle and energy, strength of spin-splitting and spin-orbit coupling, etc.
Transmittance and polarization periodically change with length of the junction. The period is determined by incident angle and energy. It is big when incident energy is high and incident angle is small. As the strength of spin-splitting increases, polarization and transmission periods become short. However, transmittance varies nonmonotonically, taking maximum value when p=0.15. Transmittance and spin polarization are little influenced by spin-orbit coupling. On the contrary, the effect of exchange field and spin-splitting strength is quite apparent. By tuning parameters and carrying numerical calculations, optimized system can be found. This will provide proper experimental settings to achieve spin selective output.

\section{Acknowledgements}
The work is partially supported by Basic Research fund of Nanjing University of Aeronautics and Astronautics with grant No.56XCA2405003.

\bibliographystyle{apsrev4-1}
\bibliography{DOUBLE_REFRACTION}
\end{document}